\documentclass[sigconf,balance=false,authorversion]{acmart}

\usepackage{submission,acmart-taps}
\usepackage{macros}
\usepackage{hyperref}

\usepackage{subcaption}
\usepackage{cleveref}

\copyrightyear{2023}
\acmYear{2023}
\setcopyright{acmlicensed}\acmConference[SUI '23]{The 2023 ACM Symposium
on Spatial User Interaction}{October 13--15, 2023}{Sydney, NSW, Australia}
\acmBooktitle{The 2023 ACM Symposium on Spatial User Interaction (SUI '23),
October 13--15, 2023, Sydney, NSW, Australia}
\acmPrice{15.00}
\acmDOI{10.1145/3607822.3614521}
\acmISBN{979-8-4007-0281-5/23/10}

\newcommand{\ie}{\emph{i.e.~}}
\newcommand{\eg}{\emph{e.g.~}}

\newcommand{\ea}{{\xspace}\emph{et~al.}\xspace}
\newcommand{\os}{{\xspace}OpenSCAD\xspace}

\newcommand{\sns}{{\xspace}Sketch-N-Sketch\xspace}
\newcommand{\dd}{drag-and-drop\xspace}
\newcommand{\csg}{CSG\xspace}

\newcommand{\code}[1]{\texttt{#1}\xspace}

\newcommand{\reffig}[1]{Figure~\ref{#1}\xspace}

\newcommand{\uquote}[1]{“\textit{#1}”}

\newcommand{\tech}[1]{\emph{#1}\xspace}

\newcommand{\bip}{\tech{Bidirectional Programming}}
\newcommand{\ps}{programmatic interfaces\xspace}
\newcommand{\dm}{direct manipulation\xspace}

\usepackage{tikz}

\newcounter{feature}
\renewcommand{\thefeature}{F\arabic{feature}}
\newenvironment{feature}[1]{\refstepcounter{feature}\paragraph{\thefeature.~#1}}

\begin{document}
\title{Introducing Bidirectional Programming in Constructive Solid Geometry-Based CAD}

\author{J. Felipe Gonzalez}
\orcid{0000-0002-0716-1689}
\affiliation{%
  \institution{Carleton University}
  \institution{Univ. Lille, CNRS, Inria, Centrale Lille, UMR 9189 CRIStAL}
  \postcode{F-59650}
  \city{Lille}
  \country{France}
}
\email{johannavila@cmail.carleton.ca}

\author{Danny Kieken}
\affiliation{%
  \institution{Univ. Lille, CNRS, Inria, Centrale Lille, UMR 9189 CRIStAL}
  \postcode{F-59650}
  \city{Lille}
  \country{France}
}
\email{danny.kieken@inria.fr}

\author{Thomas Pietrzak}
\orcid{0000-0002-2013-7253}
\affiliation{%
  \institution{Univ. Lille, CNRS, Inria, Centrale Lille, UMR 9189 CRIStAL}
  \postcode{F-59650}
  \city{Lille}
  \country{France}
}
\email{thomas.pietrzak@univ-lille.fr}

\author{Audrey Girouard}
\orcid{0000-0003-3223-105X}
\affiliation{%
  \institution{Carleton University}
  \city{Ottawa}
  \state{ON}
  \country{Canada}
}
\email{audrey.girouard@carleton.ca}

\author{Géry Casiez}
\orcid{0000-0003-1905-815X}
\affiliation{%
  \institution{Univ. Lille, CNRS, Inria, Centrale Lille, UMR 9189 CRIStAL}
  \postcode{F-59650}
  \city{Lille}
  \country{France}
}
\email{gery.casiez@univ-lille.fr}

\renewcommand{\shortauthors}{J. Felipe Gonzalez, {\it et al}.}

\begin{abstract}
3D Computer-Aided Design (CAD) users need to overcome several obstacles to benefit from the flexibility of programmatic interface tools. Besides the barriers of any programming language, users face challenges inherent to 3D spatial interaction. Scripting simple operations, such as moving an element in 3D space, can be significantly more challenging than performing the same task using direct manipulation. We introduce the concept of \textit{bidirectional programming} for Constructive Solid Geometry (CSG) CAD tools, informed by interviews we performed with programmatic interface users. We describe how users can navigate and edit the 3D model using direct manipulation in the view or code editing while the system ensures consistency between both spaces. We also detail a proof-of-concept implementation using a modified version of \os.
\end{abstract}



\begin{teaserfigure}
\centering \includegraphics[width=\textwidth]{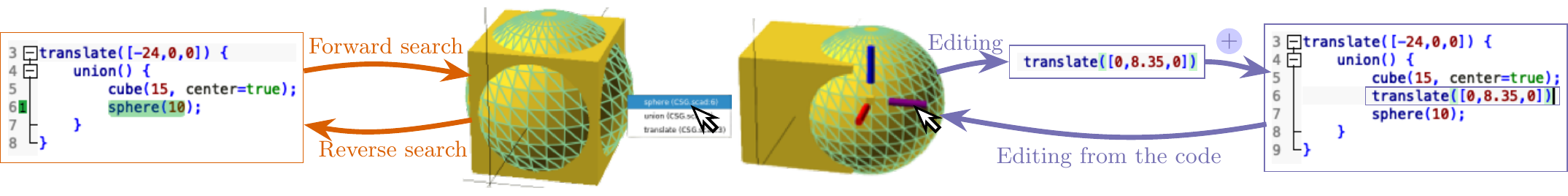}
\caption{Bidirectional Programming features implemented in \os. The system allows to navigate the code through direct manipulation in the view (reverse search) and vice versa (forward search). Also, the program enables modification of the 3D model from the view while the system updates the code coherently. }
\Description{The figure shows a code in OpenSCAD and the result after rendering: a union between a sphere and a cube. A pointer clicking on the rendered sphere part of the model is connected by green lines to another figure, the model with the edges of the sphere in green, exemplifying a selection task. Two green arrows connect this model to the code which has the sphere creation statement highlighted in green. One arrow connected from the model to the code, with the text "Reverse search", and one from the code with the text "Forward search". A title below both arrows says "Navigation." Blue lines connect the model to another figure from the rendered model with a pointer clicking in the sphere. The model with the cube and sphere union moved on one axis. A gizmo with a pointer depicts that the model was modified with a drag-and-drop action. This model is connected by a blue arrow to a piece of code, depicting a statement added to the original code to make the code coherent with the model. Above the blue arrow a title "Edition" }
\label{fig:teaser}
\end{teaserfigure}

\maketitle

\section{Introduction}

Constructive Solid Geometry (CSG) based 3D Computer-Aided Design (CAD) applications allow the creation of complex 3D objects by combining simple geometric primitives (\eg spheres and cylinders) using boolean operations (\ie union, intersection, and difference)~\cite{requicha_constructive_1977}.
CSG is widely used in programming-based software.
Unlike 3D CAD applications based on direct manipulation~\cite{shneiderman_direct_1983, the_freecad_team_freecad_2022, autodesk_inc_autocad_2022}, programmatic CSG-based CAD applications do not allow users to edit 3D objects by interacting with the mouse in the 3D view.
As a programmatic interface \cite{mathur_interactive_2020}, users have instead to iteratively edit and compile the code until they obtain the result they want.
Although users require more training to use programmatic interfaces~\cite{aish_designscript_2012}, they are relevant in the 3D printing community considering the growing number of models created from code-based designs available on websites such as Thingiverse \cite{thingiversecom_thingiverse_2022} and the large \os community \cite{openscad_openscad_2020}.

The interest in programming-based CAD relies partially on the flexibility of parametric modeling.
Designs can be easily generalized through parameters to create different customized versions by simply changing parameter values, compared to editing mesh polygons in files using the standard STL format for 3D printing \cite{library_of_congress_stl_2019}.
Consequently, parametric models are more likely to be reused than nonparametric models \cite{hofmann_greater_2018}.
For instance, the \textit{Customizer} tool from Thingiverse~\cite{thingiversecom_customizer_2022} allows storing code-based models and creating an interface for each parameter to help users adjust them to create new model versions.
Only one year after its release, Customizer was used to create about 40\% of the models on the website \cite{oehlberg_patterns_2015}.

Unfortunately, programming is a difficult task. In addition to the intrinsic difficulties of programming~\cite{shneiderman_direct_1997,preece_interaction_2015},
the most popular programmatic interface CAD programs, such as \os, present a rough static workflow, limiting the interaction to code editing and compiling dynamic.
In other words, not only programming-based 3D modeling is challenging, but the available tools do not provide features to overcome these challenges.

We examine CAD programming-based challenges and address some of them by introducing the concept \textit{bidirectional programming} into the CSG-based CAD programming field.
Bidirectional programming describes systems that allow interaction with the output of a program to update the input after defining some backward transformation always to maintain coherence between both \cite{fischer_essence_2015}.
Some GUI builders use bidirectional programming, and some research has explored it to create 2D vector graphics~\cite{hempel_sketch-n-sketch_2019}.
However, to our knowledge, the application of this concept in the context of CSG-based code editing has never been explored.
We investigate how bidirectional programming can be applied in this context.

First, we conducted semi-structured interviews with ten \os users to better understand their challenges when modeling with programmatic CAD software.
Participants reported difficulties in navigating and linking the code and the 3D view (\eg locating the code statement that creates a specific part of a model in the view). It covers understanding the contribution of specific code statements on the model and correctly relating the spatial transformation statements in the code with their effects on the model.

The insights from this study motivated the following design goals for a bidirectional programming approach to address some of the challenges found: (1) facilitating the model understanding in the view and the code through a navigating system that exploits the intrinsic relationship between both and (2) simplifying the execution of spatial transformation tasks by enabling direct manipulation interactions in a programming-based environment.
With these design goals, we present our modified version of \os (Figure \ref{fig:teaser}).
We implemented a navigation system between the code and the 3D view, allowing a better understanding of the relationships between the code and the different parts of a 3D model.
It also integrates code editing by interacting with the objects in the 3D view for operations that appear more straightforward than modifying the code.
For example, the user can translate an element of an object simply by selecting it in the 3D view with their computer mouse and adjusting its position with direct manipulation.
As a result, the values in the code that define its position would be updated, or a transformation would be added to support the translation.
Our contributions are
(1) identified challenges and practices in 3D modeling from \os users;
(2) the design of bidirectional navigation and editing features for CSG-based CAD programming;
(3) a proof-of-concept in a modified version of \os.

Our implementation and data analysis results are available as supplementary materials and at  \href{http://ns.inria.fr/loki/bp}{http://ns.inria.fr/loki/bp}.

\section{Background and Related work}

Interactive systems for content creation tools, including CAD software, can be separated into two categories.
Users can either use a Graphic User Interface (GUI) and create 3D models by manipulating objects with \dm, or they can use a code editor to describe 3D models with a program.
This section describes both approaches and details efforts to combine them. We then discuss bidirectional programming and code navigation.

\subsection{Direct manipulation}

Direct Manipulation is an interaction paradigm based on the following principles: (1) a permanent representation of the objects of interest, (2) physical actions, (3) fast, incremental, and reversible operations with an immediate and visible impact on the object of interest, and (4) progressive learning~ \cite{shneiderman_direct_1983,shneiderman_future_1982,sherugar_direct_2016}.
Most CAD software, such as Autocad \cite{autodesk_inc_autocad_2022}, Tinkercad \cite{amabilis_amabilis_2022}, or FreeCAD \cite{the_freecad_team_freecad_2022} follow the \dm approach.

This paradigm helps users to be more engaged \cite{akesson_using_2018} by reducing the cognitive resources required to understand the user interfaces \cite{shneiderman_direct_1997} and allowing them to obtain decent results with little effort \cite{aish_designscript_2012,aish_designscript_2013}.
Applying the principles of \dm also helps to achieve usability goals such as (1) constant visibility of the status of the system, (2) user control and freedom, (3) recognition rather than recall, and to some extent (4) flexibility and efficiency of use~\cite{nielsen_heuristic_1990}.
These properties make \dm more efficient than programmatic interfaces in many situations, particularly when incremental adjustments are required to reach a goal.

However, \dm has important and well-known limitations for CAD~\cite{kwon_direct_2011}.
Typically, performing repetitive tasks often results in tedious manual tasks (\eg copy-pasting an object many times) that, depending on the complexity of the output, can lead to tiresome and error-prone work.
For example, when a user needs to create multiple screw holes and place them in a model.
After creating and placing each element individually, if the user wants to modify the diameter of the hole, they need to do it one by one.
Some tasks can be challenging, such as selecting parts in models with many components due to camera occlusion and handling problems \cite{kwon_direct_2011,frohlich_history_1993}.
In addition, versioning objects, as can be done with source code, is difficult.
Further, \dm introduces ambiguity by requiring resolution heuristics from the system to interpret the user's intention when performing an action.
As a result, a similar action can have different results, making it difficult to create robust parameterized models with such software ~\cite{mathur_interactive_2020}.

\subsection{Programmatic interfaces}

Programmatic interfaces~\cite{mathur_interactive_2020} allow 3D models to be created using a textual description with code.
The code follows a formal logic structure to define objects and operations between them to create a final result.
Programming brings highly appreciated advantages over \dm systems in terms of repeatability, precision, complexity, versioning, and abstraction~\cite{yeh_craftml_2018}.
Repetitive operations, such as the screw holes example, can be expressed in a few lines of code.
In addition, the instructions are explicit, alleviating any ambiguity in the interpretation of user actions, making these interfaces a primary choice to create robust parameterized models~\cite{mathur_interactive_2020}.

However, programmatic interfaces come at the cost of learning to program in a specific language, which can be an entry barrier for users~\cite{ko_six_2004}. In addition, one of the challenges is the need to create a mental representation of a 3D object from the code \cite{qian_students_2017}.
This is even more relevant when analyzing a code written by someone else, as this task requires significant effort, even for experienced programmers.
As a consequence, 3D model designers prefer modeling tools based on \dm \cite{buehler_sharing_2015,oehlberg_patterns_2015} and most CAD applications implement \dm interfaces \cite{mathur_interactive_2020}.

\subsection{Combination of \dm and \ps}

Some efforts have improved 3D modeling by combining direct manipulation systems with programmatic interfaces.
McGuffin and Fuhrman describe a taxonomy of content creation tools based on how the output can be edited through direct manipulation or code \cite{mcguffin_categories_2020}.
We identify examples of some categories of the  CAD field.

In Blender, users can build 3D models incrementally by alternating between code statement execution and \dm editions in a \emph{Content-Oriented Programming} approach.
Users can create a mesh with direct manipulation and then execute individual code statements to edit it. This mixed solution involves coding into the direct manipulation approach. However, the code represents operations to perform on the meshes and not a representation of the 3D objects themselves.
Other programs such as FreeCAD\cite{the_freecad_team_freecad_2022} and Autodesk Maya\cite{autodesk_inc_autocad_2022} implement a \emph{Programming by Example} approach in which \dm actions are intended to teach the user how to perform these actions with code instructions, lowering the skill-requirement barrier.
Every \dm action generates the code statements in a console.
This code echoes the actions so the user can learn and execute them later with code.

Unfortunately, these approaches try to improve \dm with \ps features but do not address the latter's problems such as navigation and editing limitations.

\subsection{Bidirectional Programming}

McGuffin and Fuhrman define bidirectional programming in the context of programming interfaces that comprise both a code editor and visual content related to the set of instructions~\cite{mcguffin_categories_2020}.
Code and visual content define two different representations of the same entity.
Programming by editing the code or directly manipulating the visual content is possible.
Furthermore, any update to either representation updates the other, maintaining synchronization between the two representations.
In other words, the code always fully describes the 3D model and vice-versa.

Several parametric-based programs such as Unity\cite{technologies_unity_2023} or Rhino\-ceros\cite{rhino3d} allow selecting objects in the view and editing their properties with direct manipulation.
In such scenarios, both the description of the properties and the view representation are linked and update synchronously.
However, the approach is different from bidirectional programming because the definition of the model is not defined by code statements.
Properties define the current state of the geometry, which is different from having a code that defines all the steps to obtain the current state of a geometry.
A common example of bidirectional programming is some GUI builders with which the user can create a GUI by dragging and dropping widgets.
The corresponding coded instructions are automatically generated and can be edited to update the interface in the GUI builder window~\cite{adobe_adobe_2022}.
In a seminal work, Victor demonstrated how to generate instructions by drawing graphical elements with direct manipulation \cite{victor_drawing_2013}.
The generated instructions can then be edited to update the graphical view.
i-LaTeX~\cite{gobert_i-latex_2022} allows users to edit the content of a document from the view by adding a transitional view and improving pre-existing navigation features in \LaTeX\xspace documents. 
Mage implements direct manipulation interactions in the output of Python notebooks, allowing edits that the system reflects coherently in the code editor~\cite{kery_mage_2020}. Codelets~\cite{oney_codelets_2012} allow programmers to insert fragments of code and customize it from an editable preview so that the programmer can manipulate the intended output.

In the context of 3D CAD software, CadQuery is a script-based Python module for building parametric 3D CAD models based on B-rep representation~\cite{cadquery_cadquery_2022}. Mathur \ea~ display the output of CadQuery instructions in FreeCAD \cite{mathur_interactive_2020}.
Interaction in FreeCAD is used to select edges and relevant options, which in turn synthesize code that is inserted in the instructions.
Only operations like smoothing the edges of an object are supported.
However, it represents the first support for bidirectional programming for B-rep based CAD.
But the challenges for B-rep and CSG are different.
libfive allows users to add special variables updatable from the view~\cite{keeter_libfive_2022}.
The user can use these variables to define the model characteristics, such as the radius of a sphere.
When hovering over the surface of the model, an arrow indicates that the updatable variables control a characteristic, and the user can edit them from the view.
By dragging and dropping on the arrow, the system edits the variables to maintain coherence. 
If there are multiple variables, the system decides what variable to change.
More recently, Cascaval \ea~\cite{cascaval_differentiable_2022} present a system to edit 3D models coded with mesh-based primitives from the view.
However, it focuses on manipulating vertices parameters rather than CSG definitions and they do not propose navigation features.

\sns is a content creation tool for SVG images that leverages bidirectional programming ~\cite{hempel_sketch-n-sketch_2019}.
The system presents a programming interface with a 2D view that can be edited through \dm while the system synchronizes both.
Users can directly create basic shapes, such as rectangles or circles, in the 2D view.
Furthermore, the program places control points around the shape to control characteristics (\ie position, size, color) by clicking on them.
When a shape is created, the system inserts the code statement that creates it, including the arguments related to its characteristics.
By performing \dm, the user can update these arguments in the code from the view.
Moreover, the user can link different control points to create constraints between the figures, for example, to keep the size of two shapes equal.
As a result, a variable is created in the code instructions that are used by the different shapes and manipulated by the control points in the view.
When the user edits a characteristic controlled by multiple variable constraints, \sns uses resolution heuristics to define the best way to update the code, and the changes are propagated to other shapes using the same variables.
\sns also allows for some level of navigation in the code through the view.
When a control point is manipulated, the code editor highlights the variables that are being updated.
Finally, the view includes rectangular dashed widgets around the shapes.
The code highlights the code statements involved in the creation of the shape when the pointer hovers over the widgets.
\sns solves editing limitations for SVG using specific data structures.
Our solution focuses on 3D CSG-based modeling, which faces different challenges in model construction and 3D spatial understanding.
Typically, SVG does not support control operations such as loops used in complex cases of CSG-based CAD.
Our work highlights the importance of navigation in 3D design.
We draw inspiration from this work to implement our approach.
However, we aim not to replace programming and user control but to enhance it with helpful navigation and editing tools.

Even though modifying the instructions is the final objective, understanding how the view is connected to the source is an important non-trivial task.
Users need to exert great effort to mentally link the code with its output, which can be cumbersome~\cite{yeh_craftml_2018}.

\subsection{Navigating between the code and the view}
When the source code produces a visual output, it is possible to consider navigating the source code by interacting with its output.
For example, SyncTex allows synchronizing a \LaTeX\xspace source document with its corresponding PDF~\cite{laurens_direct_2008}. 
It is then possible to click on a sentence in the PDF viewer to jump to the corresponding line in the source document or click on a line in the source document to display the corresponding paragraph in the PDF viewer.
Similarly, modern web browsers have an inspector that allows users to navigate between elements in the web view and HTML source code.
To avoid having to switch between a code editor and the corresponding visual rendering, Gliimpse introduced a smooth in-place transition between markup code and its visual rendering~\cite{dragicevic_gliimpse_2011}.

To help understand the relationship between the code and the 3D model, the last version of \os introduced the ability to jump to the source code from the 3D preview using a right-click on a part of an object. A menu shows the different parts of the code that contribute to creating the corresponding element. However, this navigation between the code and the 3D model is not bidirectional as it is not possible to highlight which part of a model corresponds to a given line of code.
Similarly, IceSL \cite{lefebvre_icesl_2022} highlights the corresponding instructions when an object is selected.

Few examples of programming interfaces that take advantage of bidirectional programming exist, and to the best of our knowledge, this concept has not been tested in the context of CSG-based CAD.

\section{Design} \label{introducingBP}

In order to improve the programming-based CAD experience, we first needed to understand the limitations of the design process.
We interviewed ten \os users about their experiences in the design process with programmatic interface CAD programs.
Later, we determined design goals to solve some of the identified challenges, which we used in our modified version of \os.

\subsection{Initial exploration}

We experimented with the code of a dozen Thingiverse models with different complexities. We edited them to identify difficult, repetitive, or time-consuming tasks in the process.
We identified two recurring issues during the design process related to the ability to navigate the model and the ease of editing it.

The first challenge relies on the constant need to mentally connect specific parts of the code with the view and vice versa, with little to no assistance from the tool to navigate both spaces. The user needs to explore and locate in the code the exact statements for every modification based on visual inspections with no assistance. The second challenge is the need to understand the logic of the code to perform edits. The user requires a comprehensive understanding of the code statements to edit them when modifications are easy to describe on the view.

\subsection{Formative study}
We conducted semi-structured interviews to gather further knowledge about the usage of programmatic CAD systems.
We recruited 10 participants from partner laboratories and the \os Reddit channel \verb+r/openscad+ to conduct semi-structured interviews using video conferencing or in person.
All the participants had sufficient experience with \os to create and edit models.
We divided each interview into demographic questions, working observation, and bidirectional programming discussions.
The interviews lasted approximately 60 minutes on average.

We started with questions related to demographic information and previous experience with other CAD programs and programming languages. All participants were between 29 and 68 years old (average: 44.7, standard deviation: 11.4) and had over 3 years of programming experience. 8 participants had worked with direct manipulation CAD software, 3 with different programming-based CAD programs, and 2 of them only had worked with \os for 3D printing.
The participants had between 1 and more than 15 years of experience in 3D printing for leisure (4 participants), work (2 participants), or both (4 participants).
Some participants’ motivations in 3D printing were robotics, jewelry, household repairs, augmenting objects, prototyping products, and toy fabrication.

In the second part, we aimed to identify behaviors and challenges while they work in \os.
We asked the participants to bring one of their own \os models.
P2 did not provide a model, so we used one of the examples provided by \os.
The participants explained the motivation behind the model and went through the code to explain how they modeled it.
We inquired about the problems they could have and the most challenging parts of the design.
Then, we asked them to perform three simple tasks and we observed their behavior.
In the first task, we pointed at a specific part in the 3D view and asked the participant to locate the line of code that created it.
Then, in the second task, we randomly selected a different line of code and asked the participant to indicate where the contribution of that line of code was in the view.
In the last task, we asked them to perform a minor edit on the model, such as moving specific parts or resizing an element.
We asked participants to think aloud while we carefully observed the process, recurrent behaviors, and strategies.
Then we asked them about the frequency of these tasks in their typical design workflow and what difficulties they identified in \os and programmatic interfaces.

Ultimately, we shared our idea about a bidirectional system. We asked the participants if possible ideas could arise from directly interacting with the view to edit the model in a programming interface system such as \os.

\subsection{Themes}

During the interviews, we took notes of the participants' answers.
We also observed the strategies used in the second part to accomplish the tasks.  We paid particular attention to the workflow and recurrent errors committed by the participants.

One of the researchers performed an inductive thematic analysis to develop a codebook. With it, a second coder performed a deductive thematic analysis on a randomly selected interview. We calculated Cohen's kappa index to verify inter-coder reliability~\cite{braun_successful_2013} obtaining a moderated score (\(\kappa = 0.54\)).
Finally, we developed themes based on the codebook, included as supplementary material.

\subsubsection{Understanding programmatic interfaces CAD users}

Most of the participants expressed similar motivations in choosing a programmatic interface instead of a direct manipulation program for 3D modeling.
Although some participants find programming difficult, they found that coding fits their thinking, making 3D modeling accessible.
For example, P9 mentioned: \uquote{I find it hard to use what you call direct manipulation CAD programs because I am not an artist. I don't think that way, I can't sit down and make a drawing, but I can write code.}. Some of them expressed frustration using parametric direct manipulation programs as mentioned by P6: \uquote{even though I know that in Fusion360 you can work parametrically as well, but with all those purple parametric signs and all the constraints  I also get so stuck that nothing moves anymore and I have no clue why}.

As most of the participants had some programming background, they found benefits from the coding advantages, such as versioning with repositories, abstract description by including parameters to create versions of objects quickly, algorithmic description of complex geometric surfaces (\eg P2 creating fractal-based shapes or P7 creating Kumiko patterns\footnote{Kumiko is a traditional delicate and sophisticated Japanese technique of assembling wooden pieces without the use of nails \cite{yasuka_kumiko_2021}.}).
They confirmed previously known advantages of programming and expressed some problems already mentioned in the literature about direct manipulation approaches, such as the difficulty of performing repetitive tasks.

\subsubsection{Linking the code and the view}

We observed the workflow of the participants when performing navigation and editing tasks and discussed this with them after they completed it.

The participants followed two main strategies when searching for the code statements related to a part in the view. The first is a bottom-up approach. They identified the most basic module associated with the target part. Afterward, the participants tried to remember how the part is related to the surrounding elements. P6 \uquote{I always think about the basic shapes, so I know this is a cylinder in a cylinder}.
The second strategy was a top-down approach. Starting with the whole element, participants mentally split the model into subparts and select the one that contains the target part. P1: \uquote{Well, it is placed on the board, so there is a module board. Then, to create the specific parts inside...}.
The participants tried to understand how the different model objects related in both cases.
They analyzed contextual elements such as transformations, operations, or variables related to the different parts.
For instance, we asked P8 to find the code that creates one of a series of eight mounting holes in the model. They went to the variables used to define the positions and tried to understand them. P8:~\uquote{... this is the one (hole) that has a positive Y number... I called short and long side (variables). For this particular hole, it would be near 200 mm in the y-axis (short)}.
Moreover, participants use their memory exhaustively to explore the code.
P4:~\uquote{ I know that this kind of half-rounded rectangle is called a flap in my model}, P3: \uquote{I know that is the connector volume}, P7: \uquote{I know how the hex array (module) is organized in the first place}.

Participants expressed the importance of code editor features when modeling to facilitate the task. However, they stressed that the quality of \os code editor does not contribute to this goal. P7: \uquote{OpenSCAD really lacks richness in helpers how to write code.}. Many prefer to use VS Code~\cite{microsoft_visual_2023} as an external code editor because it provides highlighting and refactoring features that \os editor does not.
For instance, when a participant found a module call, to go to the module definition, they had to use the text search feature (\ie shortcut \raisebox{-3pt}{\includegraphics[width=37pt]{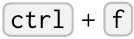}}), which could find several occurrences depending on the searched name. Other languages allow users to locate a module or function definition by holding the \raisebox{-3pt}{\includegraphics[width=20.5pt]{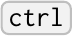}} key and clicking on the module call, for example. Participants used expressive names and generous documentation as standard good practices to facilitate code reading. Nevertheless, even the most expressive names confused them when the models had many modules and variables. Participants followed trial-and-error strategies by changing parameter values to infer their meaning.

Few participants used the search feature of \os by right-clicking on the view. We found that they could reach the code statements quickly using it, but they needed to explore many elements and read the code carefully before trying to edit it. Although the feature locates them in a line of code, it does not provide any context information that helps them understand its logic or scope. Therefore, they needed to read the code to understand it.

Participants often guessed and failed to associate a code statement with the view.
P4: \uquote{so it has to be this line here (points in the view). Wait, wait, wait ... So I think it is this line here (they change their mind)}.
After the participants thought they had reached the correct line of code of the target part, they sought confirmation.
In all cases, they sought confirmation with visual feedback.
One way to achieve this was to use a trial-and-error strategy by modifying parameter values.
They changed the location or size parameters and recompiled to check that the model changed expectedly.
Another strategy was to remove parts (\eg commenting) or create replicas (\eg create a replica of a sphere used in a \code{difference} statement).
Finally, the most common strategy was the use of modifiers.
With visual feedback, they tried to \uquote{isolate} (P5 and P6) a specific line of code contribution.
On some occasions, participants tried to edit code to get visual confirmation of changing code statements that were not doing anything in the view, such as in non-called modules of non-executed conditional, which was confusing for them.

\subsubsection{Spatial Transformation Difficulties}
Understanding the spatial dynamics in the view and connecting them with the code was a frequent problem manifested by the participants.
Having nested scopes of translation and rotation commands introduces confusion related to the coordinate system.
P6: \uquote{The most common scenario where I go wrong, and I have to literally just experiment is if I do rotate and translate together... So if you would ask me right now, rotate this in a certain direction, I would not, without testing, be able to tell you ... So for me that's always just trial and error}.

The problem manifests itself in two different ways.
On the one hand, when the design has many nested transformations, it is challenging to quickly understand what parts belong to what scopes.
P7: \uquote{you create the translate for these offsets but these translations then build above each other and then you are like wait, what one I am changing now?}.
On the other hand, it is not easy to know the center of the coordinate system and its orientation in a specific scope; programs add a gizmo to represent the 3D axis to help the user understand the reference point and orientation where they build the model.
However, once a transformation operation is applied, no visual reference guides follow those changes in the view.
Moreover, it is challenging to translate the intent of a transformation, evident in the view, into code statements.
A user may want to move an object on the X-axis.
However, if that object is defined in a nested \code{rotate}-\code{translate} scope, the user needs to calculate the corresponding transformations, which is not always easy.
This situation is straightforward in direct manipulation programs.
P2: \uquote{It's extremely easy when you want to move things in the 3D space.}.

\subsubsection{Bidirectional programming opportunities}
After we explained the concept of bidirectional programming, we asked the participants if they could think of the benefits that programming-based CAD can have by breaking the standard programming dynamic.

The participants strongly suggested moving objects directly on the view while the program updates the code coherently, adding the necessary transformations.
P5: \uquote{...just to save time, if I want to be able to move one object to the end of another object without having to go through my own model and work out how everything is, where everything goes, and result in just three static numbers...}.
Some of them went even further, suggesting that the system updates the code not with hardcoded numbers but by inferring the position where the user moves the element based on the existing variables in the code.
P4: \uquote{instead of creating the translate based on numbers it would try to find a combination of variables that lead to it}

Furthermore, the participants found it essential to be more explicit through visual cues with the relationship between the code statements and the model parts.
P2: \uquote{For example, when you hover over the text, something could happen related to the part you are hovering or selecting in the 3D view. So we should have that kind of relationship and also from the 3D view to the text.}

In addition, the participants remarked on how difficult it is to validate dimensions in \os. The structure of a model is based on the interaction of more minor elements that work together to create a model. As such, the dimensions of the entire component result from the different operations between the sizes of the smaller parts. Verifying whether the final result achieves the target dimensions is difficult in complex models. Many participants suggested a measurement tool in the programmatic interface programs.

\subsection{Design goals}

We contrasted our assumptions and the themes we found to establish two primary design goals for our approach: 1) improving the navigability of the system and 2) facilitating spatial editing.

\subsubsection{Improving the navigability of the system}\label{goal:codenav}

Typically, the user codes, and the system compiles and renders. A direct relationship exists between code statements and the different parts of the model. After visually inspecting the output, the user returns to the input to modify it. However, to do this, the system's assistance in locating the precise place in the input through existing relationships to modify the output is practically inexistent. Thus, as noted in the participants' interviews, the user needs to make this trip back from the output to the input on their own.

The system must provide interactive ways to inform users about the links between code statements and the view to facilitate navigation.
Using identifiers with visual cues, such as \os modifiers, with effective search mechanisms can significantly facilitate the design process. For instance, the user could click on a pixel in the view, and the system would show the different code statements that create it. Moreover, the user could select a code statement while the system would color the corresponding subpart in the view and highlight the code statements. This type of navigation should also be available for objects in the design that do not have a visual representation (\ie elements removed from the model in \code{intersect} and \code{difference} statements). Also, the system could provide a mechanism to visually isolate the contribution of a specific set of code statements in the view.

\subsubsection{Spatial editing}\label{goal:editing}

The participants stressed the difficulty of performing spatial transformations in programmatic interfaces due to the lack of visual assistance. Furthermore, they mentioned how easy these tasks are to perform in direct manipulation programs.

The system must provide direct manipulation actions to perform spatial transformations while keeping the code coherent. For example, the system could select a subpart in the model. The system would add visual cues to inform the current position and orientation of the subpart. The user would then perform edits through drag-and-drop mechanisms while the system adds the necessary changes to the code.

\section{Bidirectional programming for programming CSG-based CAD}
\label{sec:poc}

We created a proof-of-concept of bidirectional programming for CSG-based CAD software by patching \os because this software already has a large base of users.
Our modified version is included as supplementary material.
Before explaining the features we added to \os, we describe its overall architecture.

First, \os parses the code to create an \emph{Abstract Syntax Tree} (AST) \cite{aho_compilers_2006}, which is a structured interpretation of the \os language.
Then, it processes the AST by identifying the instantiating statements and evaluating the expressions (\eg variables, loops, functions) to create an \emph{Abstract CSG Tree} (\csg)~\cite{foley_computer_1996}.
Each node in this tree represents an element that contributes to the creation of the model and is a module instance.
The tree leaves are always primitives (\eg spheres or cylinders). Intermediate nodes can be boolean operations (\eg union), transformations (\eg translate), or groups such as control structures (\eg conditionals or loops).
Each node in this tree represents an element that contributes to the model's creation and is a module instance.
Subsequently, \os uses the \csg to compute a mesh hierarchy that contains the 3D points, normal vectors, and colors of all nodes in the \csg and stores it in a \emph{Geometric Tree}.
Finally, \os uses this tree to render the objects in the 3D view.

\begin{figure} [htbp]
\centering
\includegraphics[width=0.6\columnwidth]{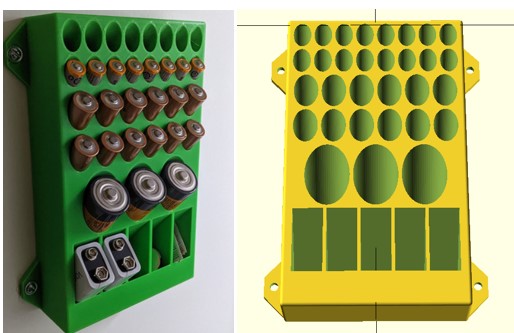}
\Description{A 3D model of a battery box retrieved from Thingiverse. On the left side the printed box. On the right side a preview of the digital model in \os.}
\caption[Caption for battery box]{A battery box model from Thingiverse \footnotemark Left: After 3D printing. Right: 3D view in \os}
\label{fig:OSProcess}
\end{figure}

We present the new features of \os in three categories: (1) reverse search navigation, (2) forward search navigation, and (3) transformations with \dm.
We illustrate these features with the 3D model depicted in \reffig{fig:OSProcess}.

\footnotetext{https://www.thingiverse.com/thing:5485266}

\subsection{Reverse search navigation}

\begin{figure}[htb]
\centering
\includegraphics[width=\columnwidth]{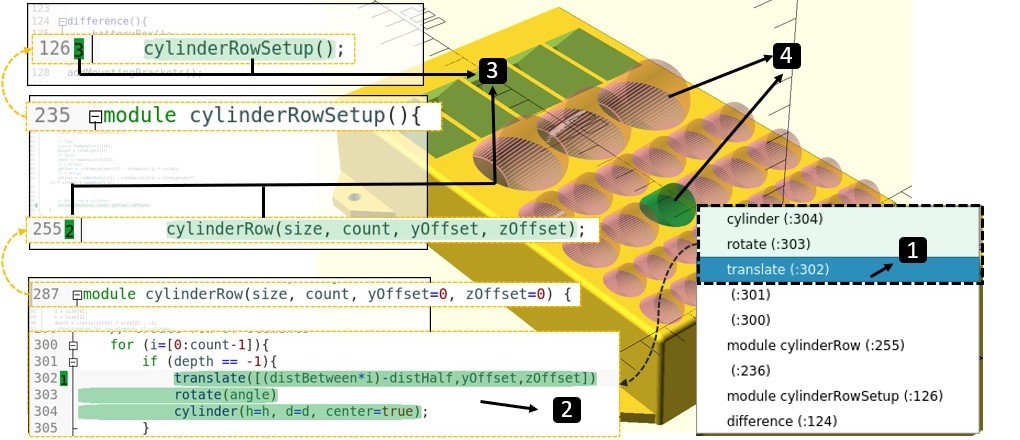}
\Description{An example of reverse search of navigation features implemented in \os. The battery box appears after one element has been selected in a reverse search. It shows the parts of the code and the model preview. The image shows how the code has been highlighted in different lines of code and the margins have been marked with a number indicating the order of call. In the model, the selected part has been colored green, and the impacted parts are colored pink.  }
\caption[Caption for reverse search features]{
Reverse navigation. \ref{feature:browsecsg}: (1) The user hovers items in the contextual menu and selects an element.
\ref{feature:highlightcode}: (2) The code of the selected element is highlighted in green. (3) Instantiating statements are also highlighted in green and marked in the margin with the call order. \ref{feature:highlight3dview}: (4) The 3D view shows ghosts of removed elements from differences, highlights the selected element in green and impacted elements in pink}
\label{figFeatures:Rev}
\end{figure}

Reverse search allows users to explore the code by interacting with the 3D model and select elements in accordance with~\ref{goal:codenav}.
For instance, the user wants to locate the line of code that creates a specific element of the model (\eg the holes for the batteries in the battery box).
Using reverse search, the user can hover or select an element in the 3D view and get visual feedback related to that element both in the 3D view and in the code editor (\reffig{figFeatures:Rev}).

\feature{Browse the CSG nodes of an element}
\label{feature:browsecsg}

In the original \os, a popup menu appears when the user right-clicks an element in the 3D view.
The items of this menu represent all the nodes in the CSG tree from the clicked element up to the root and the line number of the associated instruction in the code.

When the user clicks and \emph{selects} one of these menu items, the menu disappears, and the cursor of the code editor moves to the corresponding code statement.
If the user wants to locate the code of the other nodes, they need to click again on the same element in the view and \emph{select} a different item which breaks the navigation flow.
Also, \os does not differentiate the elements in the 3D view, and the user may be unable to make this distinction.
Hence, navigation with this selection process also exposes the user to accidentally clicking on a different element between trials, causing an error by exploring an element different from the one initially \emph{selected}.
We improved this feature by \emph{selecting} elements by hovering the pointer over the menu items.
Therefore, the user can browse the different pieces of code that created this element without closing the menu and the node references.
It helps them identify the instructions they are searching for and navigate through the code, mainly when the instructions are scattered across different parts of an extended code.

\feature{Highlight the selected element in the code editor}
\label{feature:highlightcode}

When selecting an item in the 3D view, the code displays visual cues to inform the user about the relationship between the selected element and the code.
The system recognizes two types of relationships with the target code and the subparts in the view: the \emph{target} and \emph{impacted} elements.
The \emph{target} element refers to the nodes in the branch of the \csg tree from the root to the selected node, included.
The \emph{impacted} elements refer to the other nodes of the \csg tree that were created with the same code statement of the selected part.
For instance, three elements were created with the same user-defined module.
Selecting one in the 3D view will be marked as \emph{target} while the two others will be marked as \emph{impacted}.

The code editor adds a number in the margin of the targeted nodes indicating the call order of the instruction in the call stack. It also adds a green highlight with decreasing intensity.
Further, the system highlights the code of the impacted nodes in pink. It indicates these elements will also change if the user edits the selected element.
To achieve this, the system recovers the ID of the selected element from the 3D view.
First, the system retrieves the branch of the corresponding node in the \csg tree.
Then, the system locates the lines in the code editor thanks to the reference of the AST node in each \csg node, colors the corresponding code in green, and adds the numbers indicating the call stack.
If the code of the selected element creates other elements, these other \csg nodes also have a reference to the same AST node.
Thus, the system iterates on the \csg tree to look for other elements referencing the same AST node and colors their corresponding code in pink.

\feature{Highlight the selected element in the 3D view}
\label{feature:highlight3dview}

We implemented visual feedback in the 3D view, following the logic on the code editor, to make the connection between the code and the 3D model evident.
First, the system colors in green the edges of the selected to mark it as selected.
Moreover, it colors the edges of the elements corresponding to the impacted nodes in pink.
It explicitly shows the parts that would change if the user edits this code.

Intersection and difference operations subtract the volume of elements.
Elements that produce these operations are not all clickable in the 3D view.
To address this limitation, when the selected element is one of these operations,
we draw the elements used in its creation as ghosts.
Now, the user can see and select these elements.
Ghosts are also classified as targeted or impacted and are duly colored in semi-transparent green or pink.

To implement it, when the selected element is an intersection or a difference operation, the system clones the children trees of the operation in the \csg tree. It adds to them either a semi-transparent green or transparent pink color.
Then, it adds the clones in a group and sets its position and orientation to the operation ones.

\subsection{Forward search navigation}

\begin{figure} [htb]
\centering
\includegraphics[width=\columnwidth]{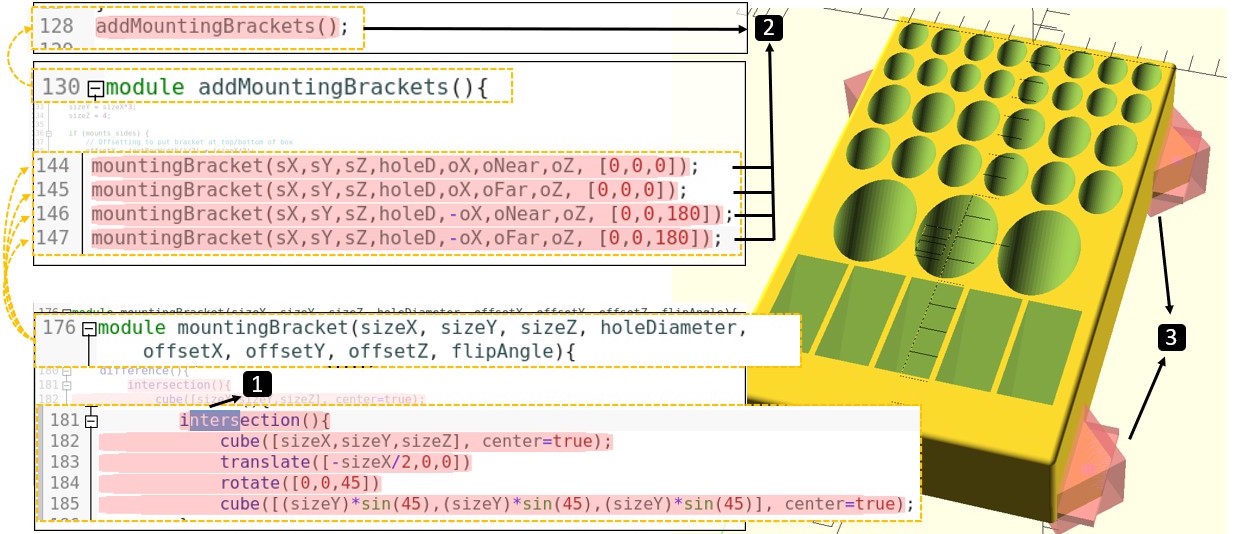}
\Description{An example of forward search of navigation features implemented in \os. The image depicts an example of a forward search. It shows the parts of the code and the model preview. The code shows a statement that has been selected. The impacted code statements have been highlighted in pink. The model has colored the impacted elements in pink    }
\caption[Caption for reverse search features part 2]{
Forward navigation. \ref{feature:fwdsearchelements}: After selecting a portion of an instantiating statement, (1) all the instance creations of the selected code are highlighted in pink, and (2) all the resulting elements are highlighted in pink in the 3D view (3).
}
\label{figFeatures:For}
\end{figure}

Forward search is the opposite of reverse search.
It allows users to explore the 3D view by interacting with the code.
For example, the user would like to understand which elements of the 3D view are created by a specific expression in the code.
Using forward search, they select this expression in the code editor, and the system highlights the impacted elements on the 3D view.
These features are illustrated in \reffig{figFeatures:For}.

\feature{Forward search in expressions corresponding to an element}
\label{feature:fwdsearchelements}

Instantiating statements contribute to creating at least one node in the \csg tree.
When the user selects two or more characters from such an expression and presses the \raisebox{-3pt}{\includegraphics[width=13pt]{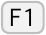}} key, the system highlights the resulting elements in the code editor and the 3D view.

The system searches all the nodes created by the selected statement. If there is only one, the program marks it as targeted; if there is more than one, it marks all nodes as impacted.
Then, the code and the model apply visual feedback as explained in \ref{feature:highlightcode} and \ref{feature:highlight3dview}.

\feature{Forward search in variables}

Variables are used in arithmetic expressions in the instruction parameters.
Therefore, modifying these variables affects the elements defined by these instructions.
When the user selects two or more characters from a variable and presses the \raisebox{-3pt}{\includegraphics[width=13pt]{F.pdf}} key, the system identifies all affected nodes and highlights in pink all the expressions in the code editor affected by this variable and the corresponding elements in the 3D view.

\subsection{Transformations with direct manipulation}

\begin{figure} [htb]
\centering
\includegraphics[width=0.8\columnwidth]{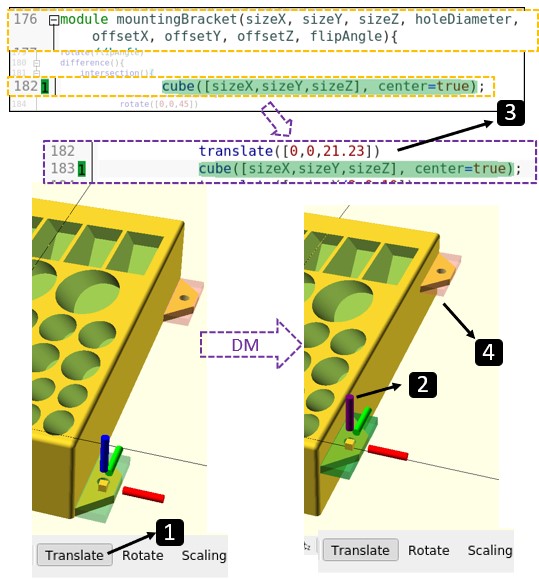}
\Description{An image showing an example of the simple edition features. The image presents lines of the code and preview before the edition and after the edition. In the code before the edition, there is an instruction to create a cube highlighted in green. After the edit, a translation instruction has been added. In the model, the selected part to be edited is colored green. The gizmo of translation appears in the center of the element. After the edition, the image shows how the part has changed its location and also how the impacted part, colored pink, has also changed. }
\caption{3D view edition. \ref{feature:translate}: (1) After selecting an element, the user enters editing mode by clicking on the translate button and a gizmo appears. (2) The user clicks and holds the z-axis and moves the pointer to the desired position. (3) The system adds a \code{translate} statement. (4) All elements impacted are also updated in the view.}
\label{fig:FeaturesSimpleEdition}
\end{figure}

CAD software typically allows users to perform transformation operations, such as translations, rotations, and scaling through \dm action on the elements on the view.
It is convenient because users can immediately validate the result and quickly set the value according to this validation.
The same task through the code requires multiple trials and errors.
We implemented similar features in \os.
For example, when users want to translate an element, they select it in the 3D view and click a \textit{translation} button in the toolbar.
A translation gizmo appears in the relative position and orientation of the selected object (\ie applying previous translation and rotation from the root to the selected object), and they can \dd one of the three axes to translate the element accordingly.
The element moves continuously and the system modifies the code simultaneously.
As moving the pointer produces large changes, the user can use the mouse wheel to make small changes of 0.1 units to achieve precise edits.
This feature is illustrated in \reffig{fig:FeaturesSimpleEdition}.

\feature{Translation from the 3D view}
\label{feature:translate}
When the user translates an element in the 3D view as described before, the system adds a \code{translate} element in the \csg tree and the code.
It adjusts the x, y, or z parameter depending on the gizmo axis the user is dragging.
The system does not add another \code{translate} element if an existing one only affects the translated element.

\feature{Rotation from the 3D view}
\label{feature:rotate}
The rotation of elements is similar to the translation described above.
The system places a rotation gizmo at the relative position and orientation of the object, which is the rotation center.
Then, the user adjusts the rotation axes through \dd.
Similarly to translations, the system only adds a \code{rotate} element if necessary; otherwise, it modifies an existing one.

\feature{Scaling from the 3D view}
\label{feature:scale}

The user can resize an element directly from the view. We added two options in the menu for this purpose: \textit{Scale} and \textit{Scale primitive}.
The user can perform the \textit{Scale} option with any selected part. If it is the only child of a \code{scale} element, the system updates the parameters of this \code{scale} element.
Otherwise, the system adds a new \code{scale} element.
Likewise, the user can perform the  \textit{Scale primitive} option if the selected part is a primitive.
The system will update the instantiating parameters.

\subsection{Informal validation by example}

\begin{figure}[htb]
\centering a)\includegraphics[height=29mm]{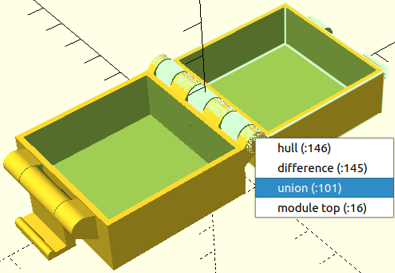}
\hspace{1mm}
b)\includegraphics[height=29mm]{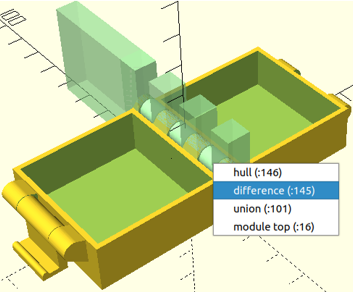}
\caption{3D View when highlighting a) the \code{union} item; b) the \code{difference} item.}
\Description{An image showing \os 3D viewer with a buckle box model from Thingiverse, a) after an element union was selected; b) when highlighting the \code{difference} item.}
\label{figTest:exBox3}
\end{figure}

We aim to show how our system addresses the design goals.
Specifically, we demonstrate it allows the user to
(1) Navigate interactively between the code and the 3D model making explicit the relationship between them, including removed elements from \code{difference} and \code{intersection} operations.
(2) Isolate the contribution of specific code statements.
(3) Perform spatial edits on the model without the need to fully understand the code.
We explored 11 models on Thingiverse under the “Popular Last 30 Days” and “Customizable” filters.
These models have on average 195 lines of code (sd 113). We performed modifications requested by Thingiverse users in the comments section of the models.
We describe below the case of a buckle box\footnote{https://www.thingiverse.com/thing:82620}~(\reffig{figTest:exBox3}), which has two parts linked by a hinge.

First, we explored how the hinge structure is built (1).
With one click on the hinges, the system displayed the menu of involved elements.
We hovered the pointer on the elements of this menu to explore the structure of this element.
The module \code{top} defines a \code{union} between a part of the box and a part of the hinge (\reffig{figTest:exBox3}).
The part of the hinge part is created by a \code{difference} statement between a \code{hull}and a set of \code{cubes}.
When hovering over the \code{difference} statement we can observe the parts used to remove some volume of the hull, represented by transparent green \code{cubes}.

\begin{figure}[htb]
\centering\includegraphics[width=\columnwidth]{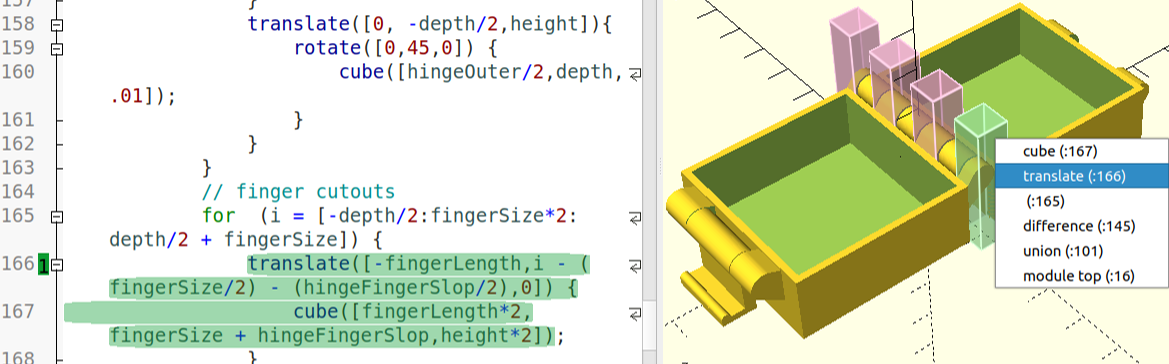}
\caption{After right-clicking on one of the transparency, code editor and 3D view highlighting the \code{translate} item.}
\Description{An image showing \os 3D viewer with a buckle box model from Thingiverse, after a translate statement was selected}
\label{figTest:exBox6}
\end{figure}

By right-clicking on one of the \code{cubes}, we repeated the navigation exercise.
When selecting the subsequent \code{translate} statement~(\reffig{figTest:exBox6}), we quickly saw that all \code{cubes} were created by the same statement when the system colors pink some of the elements in the view.
We confirmed this in the highlighted code which showed the statement inside a loop structure.
With two clicks, we could picture the code structure of a module of 84 lines of code of the model.

\begin{figure} [htbp]
\centering
\includegraphics[width=\columnwidth]{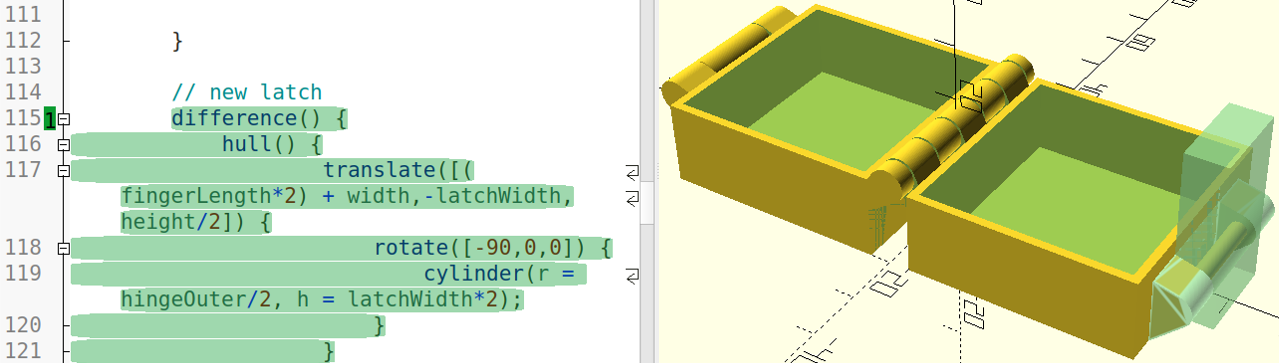}
\Description{An image showing that after placing the cursor in a code statement and pressing F1, the system highlights both, the part in the view and the statement in the code editor accordingly}
\caption{By placing the cursor on a code statement and performing a forward search, the system highlights the code and the 3D models consistently.}
\label{fixExampleBox2}
\end{figure}

Then we looked at the code to understand its logic (2).
For example, we look around further in the code and find a comment indicating the start of the description of a “new latch”.
By placing the cursor on the first code statement and pressing \raisebox{-3pt}{\includegraphics[width=13.5pt]{F.pdf}}, we could see the complete scope of the code by the highlighted text and isolate its contribution on the view with the highlighted elements and added transparencies~(\reffig{fixExampleBox2}).
Then we clicked on the transparencies on the view to observe each part individually in the code.

\begin{figure}[htb]
\centering a)\includegraphics[width=.7\columnwidth]{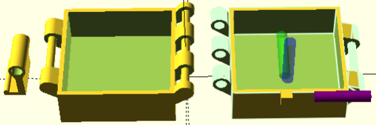}

b) \includegraphics[width=.7\columnwidth]{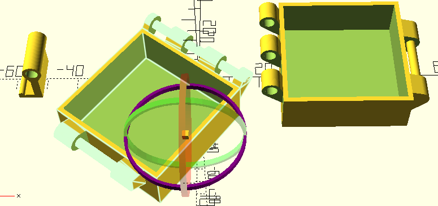}
\caption{Editing the model with a direct manipulation a) \code{translate} and b) \code{rotate} transformations on the view.}
\Description{An image showing OpenSCAD 3D viewer with a buckle box model from Thingiverse, performing a translate and a rotate transformation with direct manipulation on the viewer}
\label{figTest:exBoxT1}
\end{figure}

Last, we checked that we could perform spatial edits in the model through the 3D view (3).
For example, we aimed to perform an operation that some of the participants mentioned.
Once they finish the model, they often reorganize it to print it in an efficient way.
We further explored and realized that the model defines 3 parts.
We then selected each of these parts and performed spatial operations directly on the view to place the different parts to print them.
We started by translating and rotating only with direct manipulation actions the view
(\reffig{figTest:exBoxT1}).

\begin{figure}[htb]
\centering\includegraphics[width=\columnwidth]{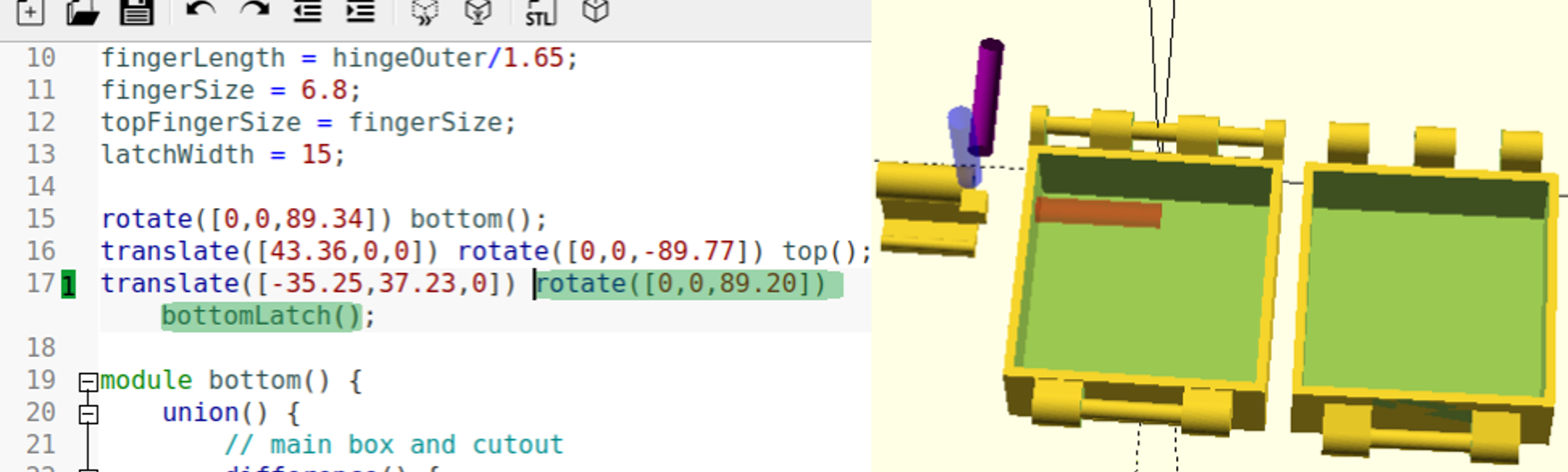}
\caption{The code updates accordingly to the spatial transformation performed on the view}
\Description{An image showing OpenSCAD 3D viewer with a buckle box model from Thingiverse and the updated code accordingly to the transformation performed}
\label{figTest:exBoxT3}
\end{figure}

We moved the parts in a satisfactory position and orientation while the system updated the code accordingly ~(\reffig{figTest:exBoxT3}).
We adjusted imprecise values from the spatial operations in the code. 
We reorganize easily the elements from the view without needing to calculate the exact angles and identify the axis.

\section{Discussion}

We discuss the results of our work, the findings of our formative study, the potential of bidirectional programming to solve problems that programming-based CAD users face, and the challenges remaining that we will address in future work.

\paragraph{Importance of programmatic interfaces.}
Previous works~\cite{mathur_interactive_2020,yeh_craftml_2018,hempel_sketch-n-sketch_2019} highlight the importance of programming in design due to its flexibility, precision, and potential in complex tasks.
In addition to confirming these technical aspects, we also found that programming-based CAD facilitates access to 3D design for programmers who cannot or do not want to use direct manipulation programs.
Our interviews show that this public includes not only developers but also engineers of different fields with math-oriented academic backgrounds.
These results suggest that there is an opportunity for HCI research to understand better in-depth and assist this group of designers in the whole process of personal fabrication.

\paragraph{Supporting the process of exploring a 3D model.}
The code helps to describe the model, whereas the view creates a visual representation that helps validate and identify errors.
If users need to edit the model based on a visual inspection of the view, they must inspect the code, analyze it, locate a specific statement, and edit it coherently.
Users need to follow a similar process if they want to reuse and adapt models to new projects.
Making this link mentally can be challenging not only because there is no trivial transformation from the view to the code but also because current tools do not facilitate this task.
Indeed, our formative study revealed that users do not know an easy way to do it.
Participants showed different strategies to achieve it that required a significant effort to understand the code followed by a visual confirmation (\eg removing statements or using modifiers).
None of them could quickly locate a code statement based on the view without studying the code.

Our solution comprehensively addresses this limitation in programmatic CAD software.
It allows users to visually and quickly understand the structure of the code-based models, solving specific problems found in the formative study: difficulty in finding a code statement that creates a part in the view, isolating the contribution of a code statement in the view, understanding the code structure when a code statement creates multiple parts (\eg a module called multiple times inside a loop), and lacking visual representation of removed objects in \code{intersect} and \code{difference} operations.

With the \emph{reverse} and \emph{forward search}, users can navigate between the code and the view back and forth.
They can identify a part in the view by clicking on it and locating the corresponding code statements involved in the text editor without studying the code while the system coherently highlights the part in the view and the corresponding code statements.
This is also helpful when exploring models from other authors, as expressed in the interviews by P6: \uquote{In other people's code, even trying to figure out what I need to isolate to see where it fits can take a lot of time}.
The system creates a visual representation of removed objects so that navigation features can be used on them.
These features improve the system of placing debug modifiers on the code, a solution that programming-based 3D CAD modelers repetitively use when designing.

Our navigation system is novel compared to existing alternatives.
IceSL~\cite{lefebvre_icesl_2022} highlights code when hovering the pointer over the model in the view, but it does not show the call hierarchy that contributed to its creation and the user cannot navigate or differentiate these instructions in the view or the code.
Sketch-N-sketch~\cite{hempel_sketch-n-sketch_2019} presents a kind of \emph{reverse search}, mainly based on the recognition of the influence of variables on the selected object rather than the instantiating statements.
Moreover, 2D SVG differs from CSG, which builds objects by operating on them, creating a tree data structure that can benefit more from our navigation system.
Finally, none of the previous work presents a \emph{forward search} feature.

We have identified non-solved challenges in our modified version of \os.
We noticed that different nodes of the code statement produce the same visual feedback when selecting elements in the view with our navigation features.
For instance, when there is a sphere inside a translation inside a rotation statement,
selecting any of these three nodes will color the sphere identically.
The users do not have a way to see on the 3D view the difference between code statements in these scenarios.
In our future work, we will investigate visual feedback for spatial transformation statements.

\paragraph{Spatial transformations.}
Spatial understanding and transformation are difficult tasks for 3D CAD modelers~\cite{mathur_interactive_2020}.
Our study showed that it can be even more challenging for programming-based CAD users because they have to transform a visual location and orientation into written operations with no visual representation of the relative coordinate system of the parts in the view.
Moreover, they do not perceive immediate and incremental feedback when modifying.
Thus, performing spatial transformation extensively requires a trial-and-error strategy.
Our approach solves this problem by placing interactive widgets on the axis of the coordinate system of the model parts.
Users can understand the relative position and orientation of the parts and edit them directly in the view.

However, spatial transformations in bidirectional programming have limitations.
Users typically use variables and arithmetic expressions in transformations to create constraints between parts of their models.
Therefore, there are several ways to modify an existing transformation: either changing the value of a variable or even changing the arithmetic expression.
\sns~\cite{hempel_sketch-n-sketch_2019} addresses this problem for SVG models by making choices with heuristics.
However, our interviews show that programming-based CAD users would like to control the model they design and make precise and deterministic modifications.
Non-code-based parametric programs, such as FreeCAD, use a \textit{constraint solver} to compute solutions that fulfill all the constraints and choose one of them when there are several possibilities.
Therefore, instead of making decisions on behalf of the user, we would like to give them control over which variables they would like to change or not.

\section{Conclusion}

We presented an adaptation of the concept of \bip to CSG-based CAD systems.
Users can browse and edit 3D models from both their programmatic description and the 3D view with \dm
We conducted semi-structured interviews with \os users to identify the way they 3D design and identify difficulties and limitations.
As a result, we identified reasons for participants' preferences in using \os rather than \dm-based CAD.
We also found that code analysis in 3D modeling requires either a fine knowledge of the code and the model, or trial-and-error procedures to navigate the code and perform edits.
Moreover, we noticed that participants struggled to perform spatial transformations, in particular when they are combined.
We explained to the participants the concept of bidirectional programming and they expressed interest in it and mentioned situations in which they would find it useful.
Then, we proposed design goals based on this study and described the features of a proof-of-concept implementation based on \os that implements them.
We describe an informal validation with a detailed walkthrough that illustrates how the new features help with the current difficulties we observed among the participants of our initial study.
Finally, we discuss the strengths and limitations of our work, as well as future work.

%

\bibliographystyle{ACM-Reference-Format}
\bibliography{references.bib}

\appendix

\section*{APPENDIX}
\setcounter{section}{1}

We describe the base questionnaire used in the semi-structured interviews.

\textit{Q1	Do you have experience with OpenSCAD? Do you have at least an advanced beginner level (you are capable of creating designs and understanding the code of a model)?}

\textit{Q2	What is your gender?}

\textit{Q3	How old are you?}

\textit{Q4	What is your academic background?}

\textit{Q5	What is your current job?}

At this point, we clarify to the participants the terms of direct manipulation and programmatic interface CAD for possible posterior discussions.

\textit{There are two main groups of CAD programs. There are direct manipulation programs in which you can directly edit the model in the view by actions with a pointer such as a drag-and-drop (i.e. Blender, AutoCAD, TinkerCAD).  Also, there are programmatic interfaces (or code-based) programs where you have to code to describe the model, then press a button to compile, and then have the result in a view, like in OpenSCAD.}

\textit{Q6	What direct manipulation CAD programs have you used before and what is your experience with each?  }

\textit{Q7	What is your skill level in OpenSCAD?}

\textit{Q8	What was your motivation to learn/use OpenSCAD specifically?}

\textit{Q9	Let's talk about the last three objects you 3D printed.}

\textit{Q10	Would you say that, in general, you 3D print for the motivations mentioned before, or are there other main reasons?}

\textit{Q11	Specifically, in the model design part, what is the most difficult and most time-consuming part?}

\textit{Q12	If different from the previous answers, specifically in OpenSCAD, what is the most difficult and time consuming part?}

\textit{Q13	Did you bring some of your previous projects in OpenSCAD?}

\textit{Q14	I will ask you to localize the specific lines of code that create a part of the model. Please say aloud the thinking process you follow to find it. Is this a task you normally do when designing, looking for a specific part of the code based on the view? What is the hardest part of doing it? }

\textit{Q15	I will ask you to localize the specific contribution of a code statement in the view. Please say aloud your thinking process you follow. Is this a task you normally do when designing (localize a part in the view based on a line of code)? What is the hardest part of doing it?}

\textit{Q16	I will ask you to perform a small edit and say aloud your thinking process while you do it.   What is the hardest part of the process of performing an edit in a pre-existing model? What edits require recompiling several times? How often do you mistake the argument in the statement you need to edit?  Why do you think this happens? }

\textit{Q17	Do you think you normally perform the previous tasks when you edit a pre-existing model? i.e. Find a part of the code based on an inspection of the view and find a specific location in the view that is created by a given code statement. }

\textit{Q18	How difficult is it to find them?}

\textit{Q19	In OpenSCAD (and programmatic interfaces), how easily can you link the output in the view with the code?}

\textit{Q20	What would you say is the best of OpenSCAD and the worst? What would you say is the best of programming-based CAD and the worst? }

\textit{Q21	What would you say are the advantages and disadvantages of direct manipulation programs and programmatic interfaces programs like OpenSCAD? }

\textit{Q22	I am working on a technology called Bidirectional programming in CAD software. In this approach, the model will be described by code, just like you do it in OpenSCAD. However, the idea is that you can modify the code using different means than the code editor. Specifically, the idea is that you can perform direct manipulation interactions, like the drag and drop you have in direct manipulation programs, while the system transforms those commands into code edits. The objective is to bring the strengths of direct manipulation tools and combine them with the flexibility of programming. According to this idea, what do you think would be useful to bring from direct manipulation tools to programming-based CAD such as \os.}

\end{document}